\documentclass[preprint,aps]{revtex4}
\usepackage{graphicx}
\usepackage{amssymb}
\usepackage{epstopdf}
\usepackage{amsmath}
\usepackage{caption,keyval,subfig}
\usepackage{listings,textcomp}
\usepackage[usenames, dvipsnames]{color}
\usepackage[ pdftex, plainpages = false, pdfpagelabels, 
                 pdfpagelayout = useoutlines,
                 bookmarks,
                 bookmarksopen = true,
                 bookmarksnumbered = true,
                 breaklinks = true,
                 linktocpage=all,
                 pagebackref=false,
                 colorlinks = true,
                 linkcolor = BrickRed,
                 urlcolor  = blue,
                 citecolor = BrickRed,
                 anchorcolor = green,
                 hyperindex = true,
                 hyperfigures
                 ]{hyperref}
                 
\newcounter{figref}

\begin{document}

\title{\href{http://www.necsi.edu/research/multiscale/paircxprofile/}{Computationally Tractable Pairwise Complexity Profile}}
\author{Yavni Bar-\!Yam}
\author{Dion Harmon}
\author{\href{http://necsi.edu/faculty/bar-yam.html}{Yaneer Bar-\!Yam}}
\affiliation{\href{http://www.necsi.edu}{New England Complex Systems Institute} \\ 
238 Main St. Suite 319 Cambridge MA 02142, USA \vspace{2ex}}

\begin{abstract}Quantifying the complexity of systems consisting of many interacting parts has been an important challenge in the field of complex systems in both abstract and applied contexts. One approach, the complexity profile, is a measure of the information to describe a system as a function of the scale at which it is observed. We present a new formulation of the complexity profile, which expands its possible application to high-dimensional real-world and mathematically defined systems. The new method is constructed from the pairwise dependencies between components of the system. The pairwise approach may serve as both a formulation in its own right and a computationally feasible approximation to the original complexity profile. We compare it to the original complexity profile by giving cases where they are equivalent, proving properties common to both methods, and demonstrating where they differ. Both formulations satisfy linear superposition for unrelated systems and conservation of total degrees of freedom (sum rule). The new pairwise formulation is also a monotonically non-increasing function of scale.  Furthermore, we show that the new formulation defines a class of related complexity profile functions for a given system, demonstrating the generality of the formalism.
\end{abstract}
\maketitle

There have been various approaches to quantify the complexity of systems in a way that is both rigorous and intuitive. The origins of these efforts are rooted in Shannon's entropy \cite{Shannon} and Kolmogorov's algorithmic complexity \cite{Kolmogorov_art}. These two measures have been modified and applied in diverse contexts \cite{Ashby,Lempel1976,Rissanen1986}. However, a disadvantage of these measures is that, being decreasing functions of order, they characterize high entropy systems---usually systems at equilibrium---as highly complex systems. On the other hand, reversing this result by defining complexity as an increasing function of order, as suggested in other work \cite{Schrodinger1944}, identifies ordered systems as highly complex. Neither equilibrium nor ordered systems match our intuitive notion of high complexity \cite{Huberman1986,Grassberger1986,Wackerbauer1994,Shiner1999}.

The literature responding to this dilemma has typically redefined complexity as a convex function of order, so that the extremes of order and randomness are both low complexity \cite{Shiner1999,LopezRuiz1995,Lloyd1988,Wackerbauer1994,Feldman1998,Gershenson2012}. One approach has not sought a single number to quantify complexity and instead argues that complexity is inherently a function of the scale of observation of the system \cite{ms_c/e,dynamics}. Because the complexity of systems lies in the emergence of behaviors from fine scale to large scale, this body of work argues that the meaningful measure of complexity represents the degree of independence among the subdivisions of the system (the disorder) as a function of the size of the subdivisions whose interdependence is considered (scale of aggregation of the observation). Because the presence of actions at one scale depends on order (coordination) on smaller scales, this measure represents the needed balance between order and disorder in a specific and meaningful way.

The dependency of complexity on scale has been termed the ``complexity profile.'' It has been applied to a variety of real world situations \cite{dynamics,ms_health,ms_military,MTW}. An expression for the profile in terms of system state probabilities has been provided and justified \cite{ms_c/e}, and it has been evaluated for a number of specific models \cite{ms_c/e,corr_gaussians,ferromagnet,chaotic_channel}. Progress is also being made in embedding the complexity profile in a theory of structure \cite{BenandBlake}. For the general case, however, this expression has the computational disadvantage of requiring factorial time to compute. In this work, we present an alternative formulation of the complexity profile, which (a) requires only polynomial time to compute, (b) satisfies key properties of the previous formulation, (c) gives the same result as the original formulation for many cases, (d) provides an approximation of the original formulation where it differs, (e) is always positive and non-increasing with increasing scale, which matches intuitive (though not necessarily valid) notions of complexity, and (f) captures some, but not all, of the high-order dependencies captured by the original formulation.

Our ``pairwise complexity profile'' is computed from only the mutual information values of each pair of components, while the original formulation requires the mutual information of all possible subsets of components. Thus, we capture those dependencies in the system that are manifest in the pairwise couplings. Such dependencies are representative of weak (and of null) emergence, but not strong emergence, while the original complexity profile also captures strong emergent dependencies \cite{strong_emerge}. The use of pairwise couplings is conceptually similar to the use of a network model or of pairwise interaction models in statistical physics. However, systems modeled by such methods may still exhibit higher-order mutual information not necessarily derivable from the pairwise couplings. The utility of this new formulation depends on the specifics of the problem, as we explore in examples.

Our formulation of the pairwise complexity profile defines a class of different measures with common properties. Distinct complexity profiles result from different functions used to measure the coupling between variables---mutual information being our primary example. In important ways the pairwise complexity profile exhibits the same behavior across a broad set of coupling functions, manifesting the general power of the framework.

The paper is organized as follows. In Section \ref{derivation}, we construct the formula for the pairwise complexity profile. In Section \ref{limiting_cases}, we show that it provides the same result as the original complexity profile for the limiting cases of a totally atomized system, and a totally coherent system. In Section \ref{properties} we prove that the pairwise formulation in general satisfies two theoretical properties of the original complexity profile---superposition of independent subsystems, and conservation of total degrees of freedom---as well as one property not shared by the original complexity profile---monotonicity. In Section \ref{generality}, we discuss the robustness of those properties to changing the definition of coupling strength. In Section \ref{limitations}, we discuss phenomena not represented in the pairwise formulation, which do appear in the original complexity profile. In Section \ref{algorithm}, we bound the computational cost of calculating the pairwise complexity profile.

\section{Derivation}\label{derivation}
We consider a system comprising a set of $m$ components, each of which may assume a number of states. The state of one component depends to some extent on the states of the other components. In particular, the likelihood of observing a system state is not equal to the product of the probabilities of the relevant component states. Different system structures are therefore reflected in distinct probability distributions of the states of the whole system. {In some contexts, for example, one may infer the probability distribution from a collection of systems as a proxy for the ensemble of possible states of a single structure, e.g. a population of biological cells. In other contexts, one may employ the ergodic hypothesis to infer the probability distribution from states of a system sampled over time, e.g. financial market data.}

We can calculate the mutal information between any two components, to obtain a measure of their ``coupling'', a value between $0$ and $1$. 
All of the couplings make up a symmetric $m\times m$ matrix, $A$.
For simplicity, in this paper we assume that each variable has the same probability distribution of states when considered independently from the others. In the examples we provide below, they are random binary variables, equally likely to be $0$ or $1$. These assumptions were also used in the mathematical development of the original complexity profile \cite{ms_c/e}.

We construct the pairwise complexity profile from a ``variable's eye view'' of the system, in order to explain its structure. First, we establish what scale and complexity mean for an individual variable or component of the system. Intuitively, for a given variable, the scale of its action is the number of variables it coordinates with to perform the action. The complexity, or the amount of information shared by that group of variables, is given by the mutual information between those variables, which is by definition their coupling. {Thus, the size of the coordinated group depends on the degree of coordination required. Motivated by this intuition, we define the ``scale of action'', $k_i(p)$, of the $i$th variable as a function of the coupling threshold $p$---the amount of coordination necessary for that action.} Explicitly, $k_i(p)$ is the number of variables coupled with the $i$th variable with coupling strength $p$ or greater:
\begin{equation}
k_i(p)=\big|\{j\in [1,m]\mid A_{ij}\geq p\}\big|
\label{k_i}
\end{equation}
where $\big|\mathcal S \big|$ indicates the number of elements in set $\mathcal S$. For $p$ a real number between 0 and 1,  $k_i$ is an integer between 1 and $m$. $k_i$ cannot be less than 1 because even for $p=1$, the highest threshold of coupling, each variable correlates with at least one variable---itself. For $p>1$, $k_i(p)=0$. Note also that $k_i(p)$ is a non-increasing function of $p$, because if a pair of variables has a coupling of at least $p_1$  then it certainly also has a coupling of at least $p_2<p_1$.

{The scale of action, $k_i(p)$, is the functional inverse of our notion of the complexity profile---we are seeking complexity as a function of scale, and $k_i(p)$ is a measure of scale as a function of complexity.} Since the coordination $p$ is a measure of the shared information, it is a measure of the complexity of the task. Below, we describe the details of taking the inverse, to account for discontinuities, which follows an intuitive mirroring over the $k=p$ diagonal. Thus we let
\begin{equation}
\widetilde{C}_i(k)=p(k_i)
\label{C_twiddle_i}
\end{equation}
which is non-increasing in all cases, and zero for $k>m$.
This gives us a profile function for each variable. Our first, non-normalized ansatz for the complexity profile for the system is to sum the functions for all of the variables:
\begin{equation}
\widetilde{C}(k)=\sum_{i=1}^{m}\widetilde{C}_{i}(k)
\label{C_twiddle}
\end{equation}
However, this must be corrected for multiple counting due to adding together the variables that share information. To illustrate the multiple counting, consider a system in which there are clusters of highly coupled variables, Eq.\ (\ref{C_twiddle}) counts the information of each cluster once for each variable in it, so larger (scale) clusters will also appear to have higher complexity. To normalize appropriately, each change (decrease) in the curve defined in Eq.\ (\ref{C_twiddle}) should be divided by the scale at which that change is found. Thus, we define the pairwise complexity profile as:
\begin{equation}
C(k)=\sum_{k'=k}^{m}\frac{1}{k'}\left[\widetilde{C}(k')-\widetilde{C}(k'+1)\right]
\label{discrete}
\end{equation}
where $k$ is a positive integer. $C(k=1)$ is the Shannon entropy of the system, and for $k>m$ we know that $C(k)=\widetilde{C}(k)=0$.

\begin{figure}[tb]
\centering
\refstepcounter{figref}\label{fig:food_pert}\label{invert_fig}
\subfloat{\label{invert_pre}\includegraphics[width=0.38\textwidth]{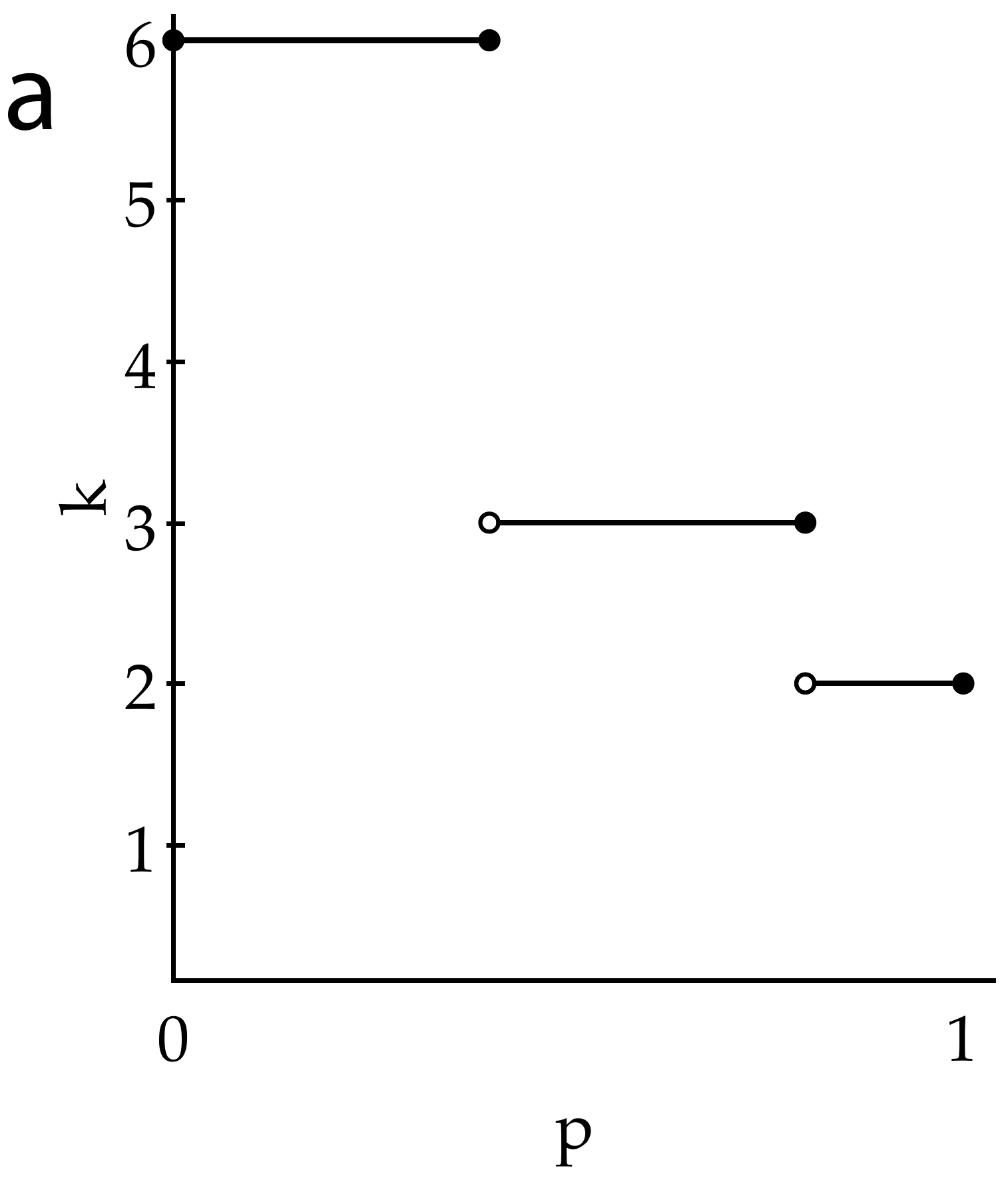}} \hfill
\subfloat{\label{invert_mid}\includegraphics[width=0.45\textwidth]{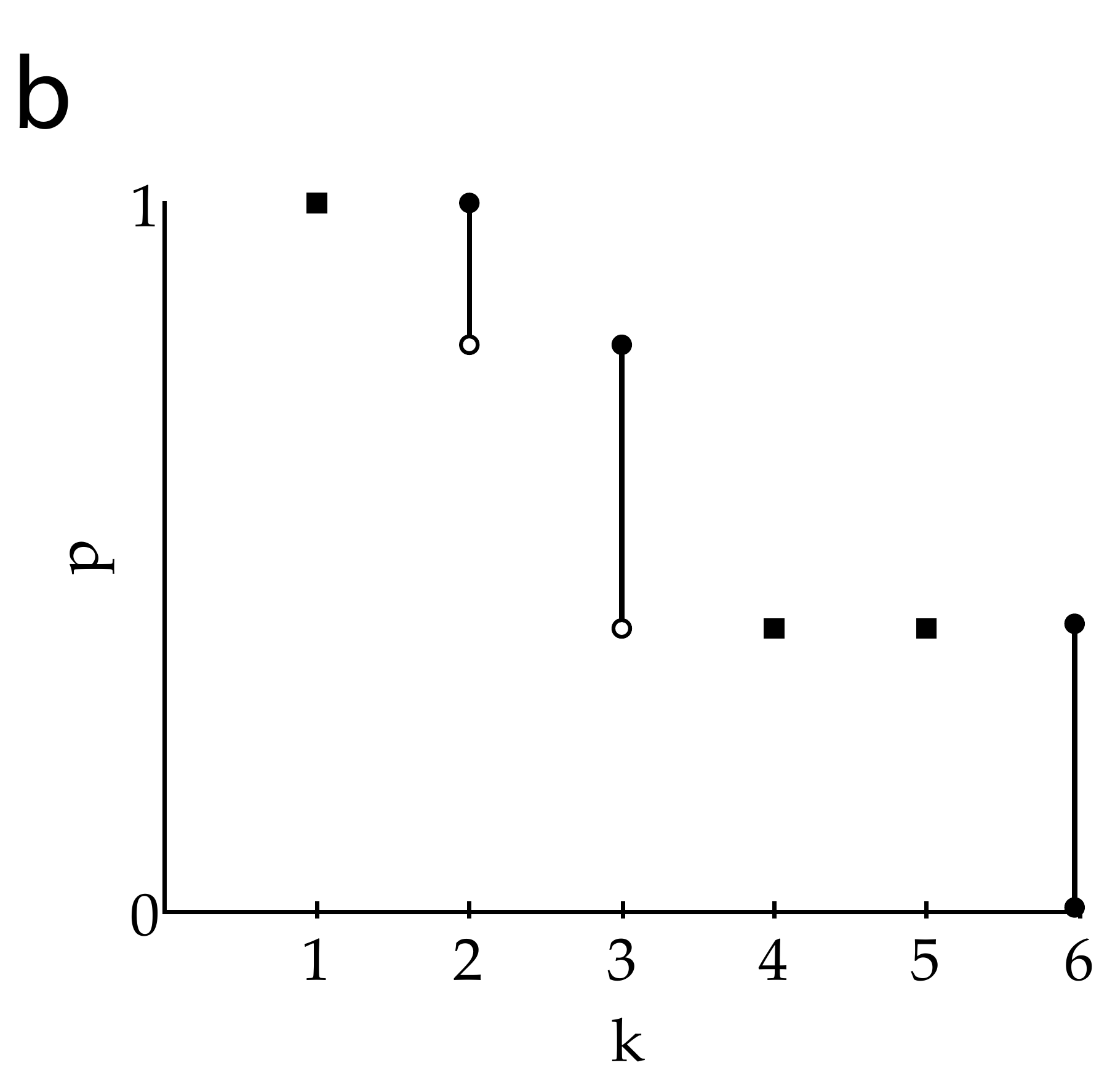}} \\
\subfloat{\label{invert_post}\includegraphics[width=0.45\textwidth]{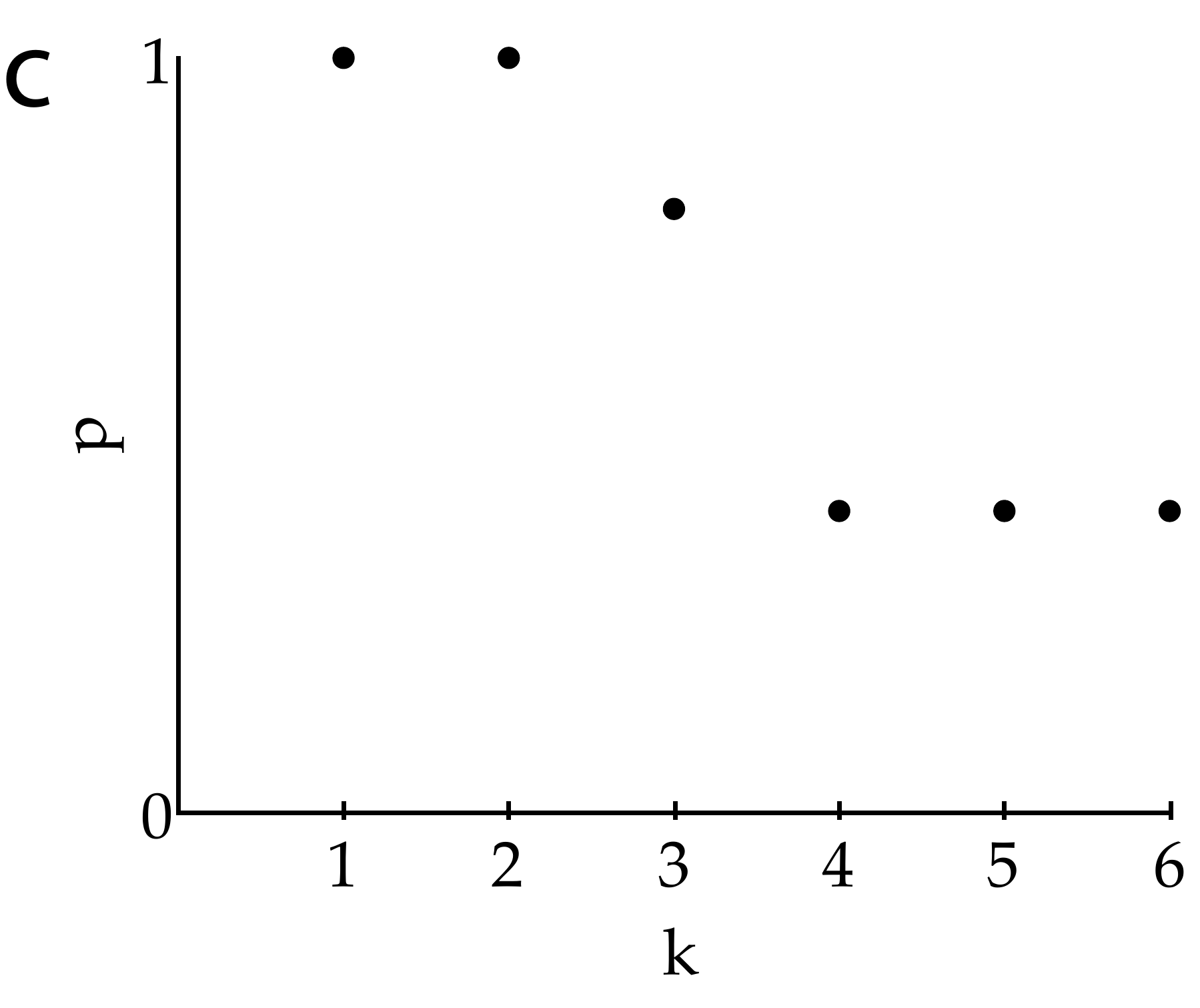}}
\caption{Illustration of the inversion process of $k_i(p)$, as described in Section \ref{derivation}: a. An example plot of a function $k_i(p)$. b. The inverse relation $\{(k,p)\}$ for the example function, augmented with additional points to fill in gaps (squares). c. The resulting function  $\widetilde{C}_i(k)=p(k)$.}
\end{figure}

For completeness, we now describe the inversion process from $k_i(p)$ to $p(k_i) = \widetilde{C}_i(k)$ (see Figure \ref{invert_fig}). Our definition for each $k_i(p)$ in Eq.\ ($\ref{k_i}$) is a mapping from the set of non-negative real numbers to a set of positive integers. We would like to represent the natural inverse function, which should be a mapping from the set of positive integers into the set of real numbers. Two issues must be resolved: more than one $p$ may map to the same $k_i$, and there are values of $p$ not mapped from any integer $k_i$. We redefine the function $k_i(p)$ as a relation described as a set of ordered pairs $\{(p,k)\}_i$. First, we fill in gaps in the $k$ axis direction with straight lines between the adjacent points. Explicitly, given a jump discontinuity at $p_0$ such that
\begin{equation}
\left\{\begin{array}{ll}
	\lim_{p\to p_0 +}k(p)=k'\\
	\lim_{p\to p_0 -}k(p)=k''
	\end{array}\right.
\end{equation}
we augment the relation $\{(p,k)\}_i$ by including the pairs with $p_0$ associated to each of the integer scales between $k'$ and $k''$, i.e. the relation now includes $\{ (p_0,k)\mid k \in \{k',\dots,k''\}\}$. Because $k$ is a discrete variable, any change in $k$ involves a jump discontinuity, but we are actually only adding points to fill in the gap when the difference is more than 1. We also include the points $\{(1,k)\mid k \in \{0,\dots,k_i(1)\}\}$, which fill in the gap to the $p$-axis, as well as the points $\{(0,k)\mid k \in \{m+1,\dots\}\}$, which extend the curve along the $k$ axis, past $k=m$. The inverse relation, $\{(k,p)\}_i$ (Figure \ref{invert_mid}) is then transformed into a function by selecting the maximum value of $p$ for each $k$, in the cases where there is more than one $p$ for a given $k$. The resulting functions (Figure \ref{invert_post}) are used as $\widetilde{C}_i(k)$ in Eq.\ (\ref{C_twiddle_i}).

\section{Limiting Cases}\label{limiting_cases}

We evaluate the pairwise complexity profile for a number of limiting cases as examples. \subsection{Ideal Gas (independent variables)} \label{ideal_gas} 
\begin{figure}[t!b]
\centering
\subfloat{\label{independent}\includegraphics[width=0.45\textwidth]{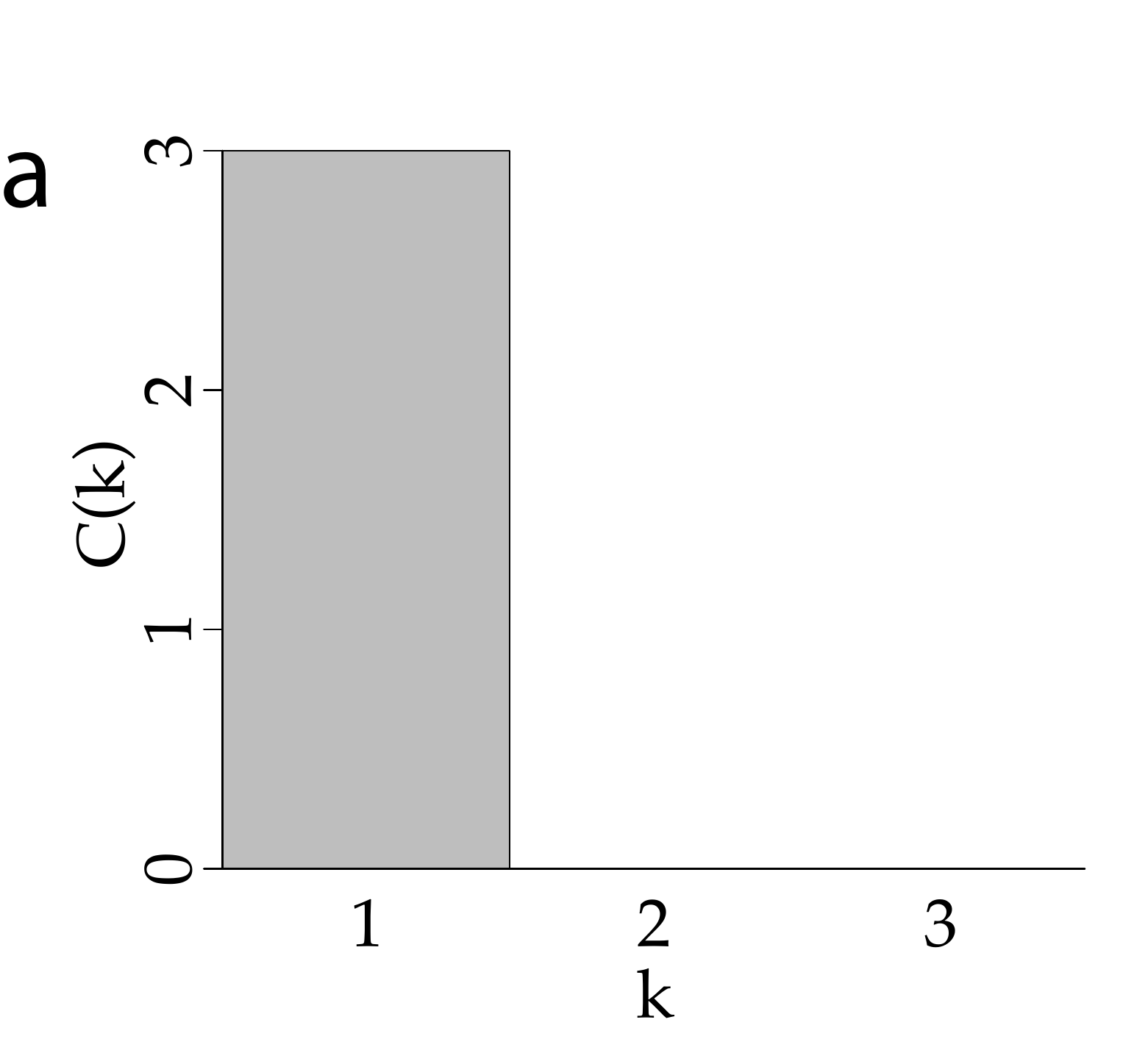}} \hfill
\subfloat{\label{coherent}\includegraphics[width=0.45\textwidth]{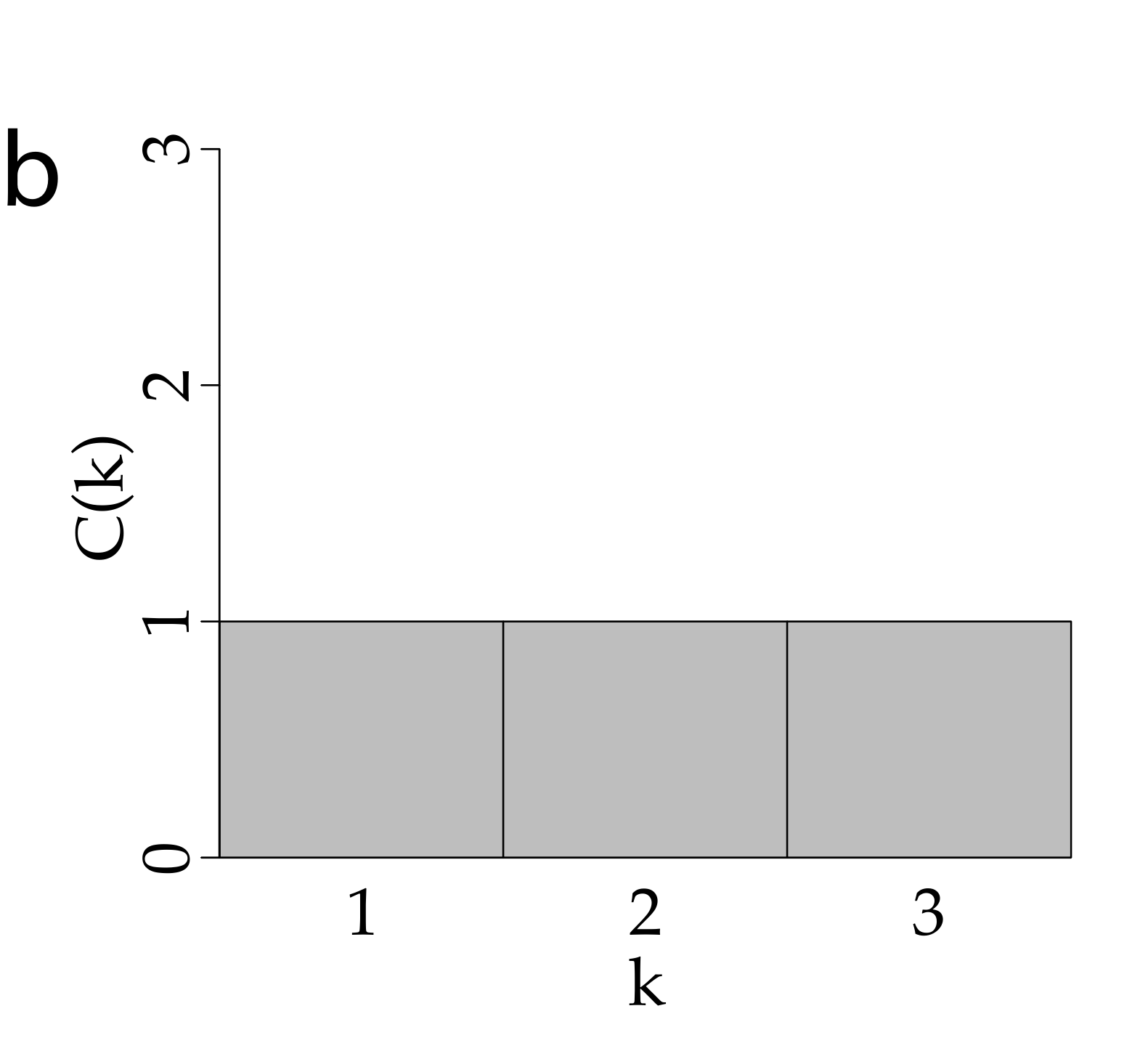}}
\caption{Complexity profiles for two systems of three variables: a. Three independent variables, the ``ideal gas'', as given by both the original and the pairwise formulations (Section \ref{ideal_gas}). As discussed in Section \ref{limitations}, this is also the pairwise complexity profile for three bits including one parity bit. b. Three coherent variables, the ``crystal'', as given by both the original and the pairwise formulations (Section \ref{crystal}).}
\end{figure}

First, we consider the case of $m$ completely independent variables, with no mutual information. The coupling matrix $A$ is the identity matrix: $A_{ij}=\delta_{ij}$, with each variable having complete coupling with itself ($A_{ii}=1$) and no coupling with any other variable ($A_{ij\neq i}=0$). Thus, for each variable, the only scale of action possible is one variable, which applies for any threshold up to and including complete coupling ($p\!=\!1$), i.e.
\begin{equation}\label{ideal_gas_k_i}
k_i(p)= \left\{ \begin{array}{ll}
         1 & \mbox{$p \leq 1$}\\
        0 & \mbox{$p>1$}
       \end{array}\right.,\forall i\in[1,m]
\end{equation}
Taking the inverse yields
\begin{equation}\label{ideal_gas_c_i_twid}
\widetilde{C}_i(k)= \left\{ \begin{array}{ll}
         1 & \mbox{$k \leq 1$}\\
        0 & \mbox{$k>1$}
       \end{array}\right.,\forall i\in[1,m]
\end{equation}
and summing over the $m$ variables gives
\begin{equation}\label{ideal_gas_c_twid}
\widetilde{C}(k)= \left\{ \begin{array}{ll}
         m & \mbox{$k \leq 1$}\\
        0 & \mbox{$k>1$}
       \end{array}\right.
\end{equation}
In this case, there is no multiple counting because any potential ``group'' of variables has only one member, so Eq.\ (\ref{ideal_gas_c_twid}) is in fact the complexity profile. If we explicitly apply the correction for multiple counting (Eq.\ \ref{discrete}), this is confirmed:
\begin{equation}
C(k=1)=\frac{1}{1}\left[m-0\right]+\sum_{k'=2}^{m}\frac{1}{k'}\left[0-0\right]=m
\end{equation}
and
\begin{equation}
C(k\neq1)=\sum_{k'=k}^{m}\frac{1}{k'}\left[0-0\right]=0
\end{equation}
so
\begin{equation}
C(k)=\left\{ \begin{array}{ll}
        m & \mbox{$k = 1$}\\
        0 & \mbox{$k>1$}
       \end{array}\right.
\end{equation}
This result is intuitive. With complete independence, at the scale of one variable, the system information is equal to the number of variables in the system. With no redundancy, the system information is zero at any higher scale. The graph of the ideal gas case for $m=3$ is shown in Figure \ref{independent}.

\subsection{Crystal (coupled variables)}\label{crystal}Next, we consider the case of $m$ completely coupled variables. The coupling matrix $A$ comprises only ones: $A_{ij}=1,\,\forall\, i,j\in[1,m]$. Thus the scale of action available to each variable is the size of the entire system, even up to and including a threshold of complete coupling ($p\!=\!1$), i.e.
\begin{equation}
k_i(p)= \left\{ \begin{array}{ll}
         m & \mbox{$p \leq 1$}\\
        0 & \mbox{$p>1$}
       \end{array}\right.,\forall i\in[1,m]
\end{equation}
Thus, the inverse functions are
\begin{equation}
\widetilde{C}_i(k)= \left\{ \begin{array}{ll}
         1 & \mbox{$k \leq m$}\\
        0 & \mbox{$k>m$}
       \end{array}\right.,\forall i\in[1,m]
\end{equation}
and summing these over the $m$ variables yields
\begin{equation}
\widetilde{C}(k)= \left\{ \begin{array}{ll}
         m & \mbox{$k \leq m$}\\
        0 & \mbox{$k>m$}
       \end{array}\right.
\end{equation}
Each variable is counting all of the other variables in its group of coordination for all scales. When we sum them, we are actually counting the same information for each of them. There is therefore no need to sum them, or, if we do, we divide the result by the size of the system. We explicitly use Eq.\ \ref{discrete}, to get this result:
\begin{equation}
C(1\leq k \leq m-1)=\sum_{k'=k}^{m-1}\frac{1}{k'}\left[m-m\right]+\frac{1}{m}\left[m-0\right]=1
\end{equation}
and
\begin{equation}
C(k=m)=\frac{1}{m}\left[m-0\right]=1
\end{equation}
and
\begin{equation}
C(k>m)=0
\end{equation}
so
\begin{equation}
C(k)=\left\{ \begin{array}{ll}
        1 & \mbox{$k \leq m$}\\
        0 & \mbox{$k>m$}
       \end{array}\right.
\end{equation}With all of the variables coordinated, there is only one bit of information at any scale. Because they are all coordinated, this information is represented at all scales up to the size of the system. The graph of the crystal case for $m=3$ is shown in Figure \ref{coherent}.

\section{Properties}\label{properties}

We prove that the pairwise complexity profile satisfies three properties: superposition of uncoupled subsystems, conservation of total degrees of freedom, and monotonicity. The first two properties have been shown to hold for the original formulation, and appear to be fundamental to the idea of the complexity profile \cite{ms_c/e,dynamics,ms_variety,sumrule}. Monotonicity, on the other hand, does not hold in general for the original formulation \cite{ms_c/e,strong_emerge,ferromagnet}.

\subsection{Superposition}
Given two uncoupled systems, $S_1$ and $S_2$, with pairwise complexity profiles $C^1(k)$ and $C^2(k)$, respectively, the complexity profile of the combined system $S=S_1 \cup S_2$ is given by $C=C^1+C^2$. The superposition principle is a generalization of the extensivity of entropy \cite{ms_c/e}.

\emph{Proof:} The coupling matrix of the combined system $S$, with elements indexed to correspond to those of $S_1$ and $S_2$ in the natural way, is the block matrix
\[ \left[ \begin{array}{cc}
A_1 & 0 \\
0 & A_2\end{array} \right]\]
where $A_1$ and $A_2$ are the $m_1\times m_1$  and $m_2\times m_2$ coupling matrices of $S_1$ and $S_2$, respectively, as individual systems. Zeros are appropriately dimensioned null blocks. The scales of action of each variable in the combined system are its scales of action in its own subsystem, i.e.
\begin{equation}
k_i(p)= \left\{ \begin{array}{ll}
         k^1_i(p) & \mbox{$0<p;\ i\in\{1,\dots,m_1\}$}\\
         k^2_{i-m_1}(p) & \mbox{$0<p;\ i\in\{m_1+1,\dots,m_1+m_2\}$}\\
       \end{array}\right.
\end{equation}
because it has no coordination with any of the variables in the other subsystem.
From the definition (Eq.\ \ref{k_i}), each of the $k^1_{i}(p)$'s are bounded by $0\leq k^1_{i}(p)\leq m_1$ and the $k^2_{i}(p)$'s are bounded by $0\leq k^2_{i}(p)\leq m_2$ and both are non-increasing. Thus, the inverse functions are
\begin{equation}
\widetilde{C}_i(k)=\left\{ \begin{array}{ll}
	\widetilde{C}^1_{i}(k) &  i\in\{1,\dots,m_1\}\\
	\widetilde{C}^2_{i-m_1}(k) &  i\in\{m_1+1,\dots,m_1+m_2\}\end{array}\right.
\end{equation}
where we know that $\widetilde{C}^1_{i}(k)=0$ for $k>m_1$ and $\widetilde{C}^2_{i}(k)=0$ for $k>m_2$. Their sum is therefore decomposable into the two subsystems: 
\begin{equation}
\widetilde{C}(k)=\sum_{i=1}^{m_1+m_2}\widetilde{C}_i(k)=\sum_{i=1}^{m_1}\widetilde{C}^1_{i}(k)+\sum_{i=m_1+1}^{m_1+m_2}\widetilde{C}^2_{i-m_1}(k)=\widetilde{C}^1(k)+\widetilde{C}^2(k)
\end{equation}
where $\widetilde{C}^1(k)=0$ for $k>m_1$ and $\widetilde{C}^2(k)=0$ for $k>m_2$ and therefore $\widetilde{C}(k)=0$ for $k>\max(m_1,m_2)$. 
When we correct for multiple counting, the decomposition into the two subsystems is maintained:
\begin{equation}\begin{array}{r@{{}={}}l}
C(k)&\displaystyle\sum_{k'=k}^{m_1+m_2}\frac{1}{k'}\left[\widetilde{C}^1(k')+\widetilde{C}^2(k')-\left(\widetilde{C}^1(k'+1)+\widetilde{C}^2(k'+1)\right)\right]\\
\noalign{\bigskip}
&\displaystyle\sum_{k'=k}^{m_1+m_2}\frac{1}{k'}\left[\widetilde{C}^1(k')-\widetilde{C}^1(k'+1)\right]+\sum_{k'=k}^{m_1+m_2}\frac{1}{k'}\left[\widetilde{C}^2(k')-\widetilde{C}^2(k'+1)\right]
\end{array}\end{equation}
and because $\widetilde{C}^1(k)=0$ for $k>m_1$ and $\widetilde{C}^2(k)=0$ for $k>m_2$, this is
\begin{equation}
C(k)=\sum_{k'=k}^{m_1}\left[\widetilde{C}^1(k')-\widetilde{C}^1(k+1)\right]+\sum_{k'=k}^{m_2}\left[\widetilde{C}^2(k')-\widetilde{C}^2(k'+1)\right]=C^1(k)+C^2(k)
\end{equation}
which demonstrates superposition.

\subsection{Sum Rule}
The area under the complexity profile curve of a system depends only on the number of variables in the system, not on the dependencies between them. Therefore, given a number of variables, choosing a structure of dependencies involves a tradeoff between information at a large scale (redundancy) and information at a smaller scale (variability) \cite{sumrule}.

\emph{Proof:} The total area under the complexity profile curve is given by
\begin{equation}
\begin{array}{r@{{}={}}l}
\displaystyle\sum_{k}C(k)&\displaystyle\sum_{k=1}^{m}\sum_{k'=k}^{m}\frac{1}{k'}\left[\widetilde{C}(k')-\widetilde{C}(k'+1)\right]\\
\noalign{\bigskip}
&\displaystyle\sum_{k=1}^{m}k\cdot\frac{1}{k}\left[\widetilde{C}(k)-\widetilde{C}(k+1)\right]\\
\noalign{\bigskip}
&\displaystyle\sum_{k=1}^{m}\left[\widetilde{C}(k)-\widetilde{C}(k+1)\right]\\
\noalign{\bigskip}
&\widetilde{C}(1)-\widetilde{C}(m+1)\\
\noalign{\bigskip}
&\widetilde{C}(1)=m
\end{array}
\label{sum_rule}
\end{equation}
 where the last equality holds because we know that at a coupling threshold of one (the maximum threshold), there will be at least one variable meeting the threshold condition for each variable (namely, itself), so $\widetilde{C}_i(1)=1$. If there are other variables identical to the $i$th variable, it may be that $k_i(1)>1$, but because of the way we defined the inverse function---in particular, filling in the gap to the $p$-axis---it is still the case that $\widetilde{C}_i(1)=1$. Therefore, from Eq.\ (\ref{C_twiddle}),
 \begin{equation}
 \widetilde{C}(1)=\sum_{i=1}^{m}1=m
 \end{equation}
The notion that the organization of dependencies, which define the structure of a system, is equivalent to a prioritization of complexity at different scales has been particularly useful in applications of the complexity profile \cite{ms_c/e,dynamics,ms_variety}.

\subsection{Monotonicity} The pairwise complexity profile is a monotonically non-increasing function of scale. This is not the case for the original complexity profile \cite{ms_c/e,strong_emerge,ferromagnet}. A complexity profile formalism that as a rule does not increase has the advantage of a certain intuitiveness. As one ``coarse grains'', i.e. increases the scale of observation, one expects a loss of information. On the other hand, the full complexity profile demonstrates that interesting phenomena such as ``frustration''---e.g. the anticorellation of three binary variables with each other---are captured by oscillatory complexity profiles. Section \ref{limitations} gives a simple example where this difference is manifest: the parity bit. We now prove that the pairwise complexity profile is monotonically non-increasing.

\emph{Proof:} The way we have defined complexity as a measure of information is such that if there is certain information present at a given scale, then that information is necessarily present at all smaller scales. From Eq.\ (\ref{k_i}), any couplings that satisfy some threshold $p_1$ will also satisfy any lesser threshold $p_2<p_1$, which means that $p_2<p_1\Rightarrow k_i(p_2)\leq k_i(p_1)$, i.e. $k_i(p)$ is a non-increasing function. Its inverse $\widetilde{C}_{i}(k)$ therefore is non-increasing as well, as is the sum of those non-increasing functions, $\widetilde{C}(k)$. From Eq.\ (\ref{discrete}),
\begin{equation}
C(k)=C(k+1)+\frac{1}{k'}\left[\widetilde{C}(k')-\widetilde{C}(k'+1)\right] \geq C(k+1)
\end{equation}
because $\widetilde{C}(k') \geq \widetilde{C}(k'+1)$. This proves that $C(k)$ is non-increasing.
 
 \section{Generality of the Complexity Profile}\label{generality}
The analysis thus far has been based on using the mutual information as the pairwise coupling. However, the properties derived are largely independent of the particular coupling function used, i.e.\ how the coupling strength is determined from the system state probabilities. This means that the properties derived in Section \ref{properties} arise from the conceptual and mathematical underpinnings of the complexity profile, rather than from the specific representation of the system. 

The constraints on the definition of the coupling function are minimal. In general, the coupling function must be normalized to vary between zero and one. For superposition to hold, ``unrelated subsystems'' must still be defined as having a coupling strength of zero for any pairs of components split between the two subsystems. For the sum rule to hold, each variable must couple to itself with a strength of one (to be precise, it is sufficient that each variable couple to \emph{some} variable with a strength of one). The mutual information has a logarithmic dependence on state probabilities, and its use in the pairwise formulation should lead to a good approximation of the original complexity profile. Another possible coupling strength function is the absolute value of correlation between variables, which has linear dependence on state probabilities.
 
The generality of the complexity profile formulation is particularly meaningful for the sum rule property. In that context, it is helpful to consider the freedom of variation in the definition of $\widetilde{C}$, rather than the coupling function. We can interpret $\widetilde{C}(k)/m$ as the average threshold of the coupling strength to imply a group of size $k$. The independence of the sum rule to different forms of this function means that the tradeoff between complexity at different scales is a valid feature of the complexity profile, regardless of how one identifies sets of elements as groups based on coupling strength thresholds.

Physically, one might interpret this in the following way. An external force of a certain magnitude will destroy the weaker internal couplings between components of a system. Stronger couplings are less susceptible to being broken than weaker ones. As the strength of the force increases more couplings are progressively broken. However, different particular dependencies between the strength of the external force and the strength of the couplings broken are possible; it may for example be linear or logarithmic. What we have shown is that the sum rule is agnostic to the particular functional form of that dependence. Groupings that disappear at one scale due to increased sensitivity to the external force always reappear at another scale, maintaining the same sum over all scales.

 \section{Differences between the original and the pairwise Complexity Profiles}
 \label{limitations}
The pairwise complexity profile (using mutual information as coupling) and the original complexity profile differ for cases in which there are large-scale constraints on the states of the system that are not derivable from the pairwise interactions of the components of the system. Such cases have been termed ``type 2 strong emergence'' \cite{strong_emerge}.
 
A simple example of such a system is that of the parity bit \cite{strong_emerge}. We have three binary variables, $S = \left\{x_1,x_2,x_3\right\}$, in which one of the variables is a parity bit for the other two, being $0$ if the sum of the other two is even, and $1$ otherwise. Mathematically, $x_3 = x_1 \oplus x_2$, where $\oplus$ is addition modulo two, the exclusive or ($XOR$) gate. This system is symmetric---any one of the three bits serves as a parity bit for the other two. In this case, the possible states of the system are 
\begin{equation}\tag{T1}\label{t1}
\begin{array}{cc}
\mbox{state} & \mbox{probability} \\
 \hline
\mathbf{000} & \mathbf{.25}  \\
001 & 0 \\
010 & 0 \\
\mathbf{011} & \mathbf{.25}   \\
100 & 0 \\
\mathbf{101} & \mathbf{.25}  \\
\mathbf{110} & \mathbf{.25} \\
111 & 0 \\
        \end{array}
\end{equation}

\begin{figure}[t!b]
\centerline{
	\includegraphics[width=105mm]{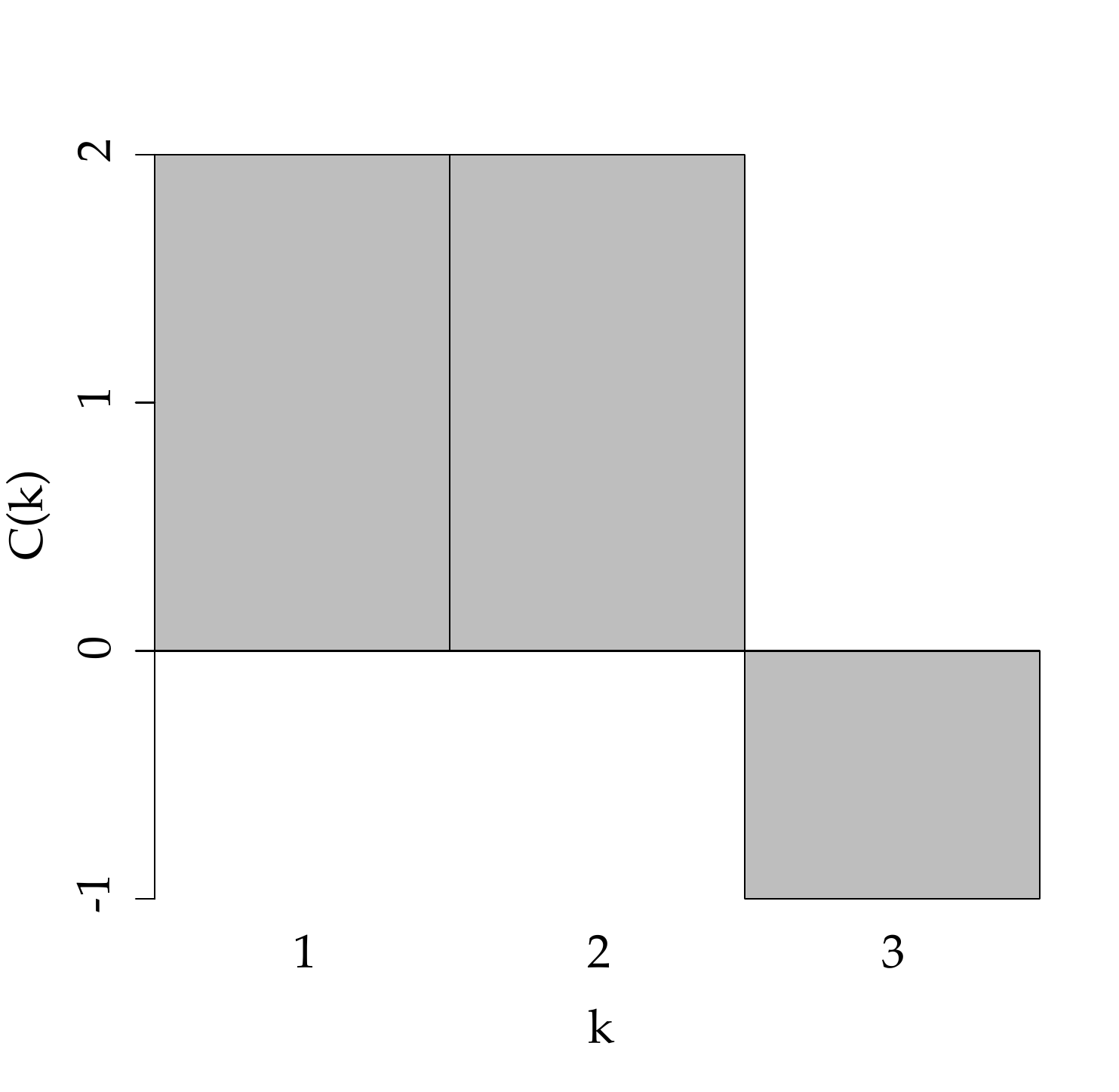}
}
\caption{Original complexity profile for three bits including one parity bit (Section \ref{limitations}). Compare to pairwise complexity profile shown in Figure \ref{independent}.}
\label{parity_exact}
\end{figure}

The probabilities of the states of each pair of variables are therefore

\begin{equation}\tag{T2}\label{t2}
\begin{array}{cc|cc|cc}
\mbox{state} & \mbox{probability} &\mbox{state} & \mbox{probability} &\mbox{state} & \mbox{probability} \\
 \hline
00\textvisiblespace & .25 & 0\textvisiblespace0 & .25 & \textvisiblespace00 & .25  \\
01\textvisiblespace & .25 & 0\textvisiblespace1 & .25 & \textvisiblespace01 & .25  \\
10\textvisiblespace & .25 & 1\textvisiblespace0 & .25 & \textvisiblespace10 & .25 \\
11\textvisiblespace & .25 & 1\textvisiblespace1 & .25 & \textvisiblespace11 & .25 \\
        \end{array}
\end{equation}
where, for example, the probability of $00\textvisiblespace$ is $P(00\textvisiblespace)=P(001)+P(000)$. The probability distributions of the states of the individual bits are
\begin{equation}\tag{T3}\label{t3}
\begin{array}{cc|cc|cc}
\mbox{state} & \mbox{probability} &\mbox{state} & \mbox{probability} &\mbox{state} & \mbox{probability} \\
 \hline
0\textvisiblespace\textvisiblespace & .5 & \textvisiblespace0\textvisiblespace & .5 & \textvisiblespace\textvisiblespace0 & .5  \\
1\textvisiblespace\textvisiblespace & .5 & \textvisiblespace1\textvisiblespace & .5 & \textvisiblespace\textvisiblespace1 & .5 \\
        \end{array}
\end{equation}
These distributions gives the complexity profile shown in Figure \ref{parity_exact}.

However, note that the probability distributions of the individual bits and of the pairs of bits (\ref{t2} and \ref{t3}) are identical to the distributions which would be derived from the case of three completely independent bits, giving the same pairwise correlation matrix,
\begin{equation}
A= \left[ \begin{array}{ccc}
        1&0&0\\
        0&1&0\\
        0&0&1
        \end{array}\right]
\end{equation}
which gives the approximated complexity profile shown in Figure \ref{independent}.

The pairwise complexity profile thus has limited applicability as an approximation to the original complexity profile in cases where large scale constraints dominate the fine-scale behavior of the system. However, when the dominant flow of information is the emergence of large scale structure from the fine-scale behavior, this approximation should be useful. We note that just as higher order correlations can result from pairwise dependencies (as in network dynamics), pairwise correlations can capture some higher-order dependencies, though not all.

\section{Computation Time}
\label{algorithm}
One of the major motivations behind developing the pairwise formulation of the complexity profile is that the existing mathematical formulation is prohibitively expensive to compute for an arbitrary high-dimensional system. The original formulation requires a calculation that involves the mutual information of every subset of components, so the computation time would grow combinatorially with the size of the system. No algorithm is known to perform the computation in time less than $\Omega(N!)$.

The pairwise complexity profile of a system of $N$ variables can be computed by an algorithm whose most time-expensive step is ordering the $N$ couplings of each of the $N$ variables.
Because efficient algorithms can sort lists of $N$ elements in $\mathcal O(N\,\log(N))$ time, the pairwise complexity profile scales as $\mathcal O(N^2\,\log(N))$.

\section{Conclusion}
We present a method of calculating a measure of complexity as a function of scale. The method is based on the concept of the complexity profile, which has proven to be useful for mathematically capturing the structure of a system of interdependent parts. The previous mathematical formulation of the complexity profile, while powerful, is too demanding computationally for many real-world systems. By offering a set of related alternative formulations, we both provide an approximation tool and demonstrate that the principles of the complexity profile, including superposition and sum conservation, are more general than a particular formulation.

\end{document}